\documentclass[acmtog]{acmart}
\acmSubmissionID{1568}

\setcopyright{acmlicensed}
\acmJournal{TOG}
\acmYear{2025} 
\acmVolume{44} 
\acmNumber{6} 
\acmArticle{184} 
\acmMonth{12}
\acmDOI{10.1145/3763312}

\usepackage{graphicx} 
\usepackage{amsmath}
\usepackage{enumitem}
\usepackage{multirow}

\newcommand{\eg}{{\textit{e.g., }}}
\newcommand{\ie}{{\textit{i.e., }}}

\newcommand{\ignore}[1]{}

\begin{document}

\title{DiffTex: Differentiable Texturing for Architectural Proxy Models}

\author{Weidan Xiong}
\email{xiongweidan@gmail.com}
\affiliation{
	\institution{CSSE, Shenzhen University}
	\country{China}	
}

\author{Yongli Wu}
\email{ngligo16@gmail.com}
\affiliation{
	\institution{CSSE, Shenzhen University}
	\country{China}	
}

\author{Bochuan Zeng}
\email{zbc8301@gmail.com}
\affiliation{
	\institution{CSSE, Shenzhen University}
	\country{China}	
}

\author{Jianwei Guo}
\email{gjianwei.000@gmail.com}
\affiliation{
	\institution{Beijing Normal University}
	\country{China}	
}

\author{Dani Lischinski}
\email{danix3d@gmail.com}
\affiliation{
	\institution{The Hebrew University of Jerusalem}
	\country{Israel}	
}

\author{Daniel Cohen-Or}
\email{cohenor@gmail.com}
\affiliation{
	\institution{Tel Aviv University}
	\country{Israel}	
}

\author{Hui Huang}
\email{hhzhiyan@gmail.com}
\authornote{Corresponding author: Hui Huang (hhzhiyan@gmail.com)}
\affiliation{
	\institution{CSSE, Shenzhen University}
	\country{China}
}

\renewcommand\shortauthors{W. Xiong, Y. Wu, B. Zeng, J. Guo, D. Lischinski, D. Cohen-Or and H. Huang}

\begin{abstract}

Simplified proxy models are commonly used to represent architectural structures, reducing storage requirements and enabling real-time rendering. However, the geometric simplifications inherent in proxies result in a loss of fine color and geometric details, making it essential for textures to compensate for the loss. Preserving the rich texture information from the original dense architectural reconstructions remains a daunting task, particularly when working with unordered RGB photographs. We propose an automated method for generating realistic texture maps for architectural proxy models at the texel level from an unordered collection of registered photographs. Our approach establishes correspondences between texels on a UV map and pixels in the input images, with each texel’s color computed as a weighted blend of associated pixel values. Using differentiable rendering, we optimize blending parameters to ensure photometric and perspective consistency, while maintaining seamless texture coherence. Experimental results demonstrate the effectiveness and robustness of our method across diverse architectural models and varying photographic conditions, enabling the creation of high-quality textures that preserve visual fidelity and structural detail.

\end{abstract}


\begin{teaserfigure}
\centering
\includegraphics[width=\linewidth]{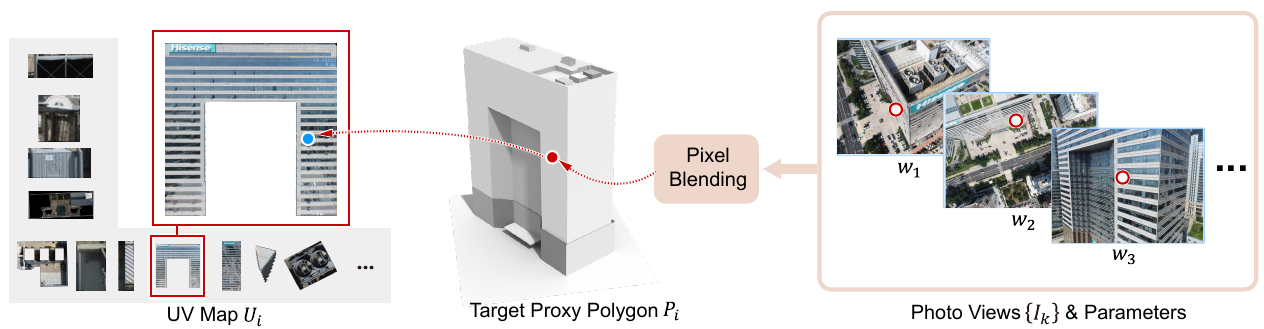}
\caption{DiffTex generates realistic texture maps for architectural proxy models. Each texel on the texture map (blue point) is defined as a blend of pixels from photo views (red circles).}
\label{fig:teaser}
\end{teaserfigure}

\maketitle

\section{Introduction}
\label{sec:intro}

The digital reconstruction of architectural models has long been a central challenge in computer graphics \cite{sinha2008interactive,garcia2013automatic,verdie2015lod,UrbanScene3D}. Advances in scanning technologies, particularly the proliferation of drone-mounted cameras, have significantly expanded the availability of high-resolution data, enabling the reconstruction of detailed models with remarkable geometric and photometric precision. However, such high-fidelity representations impose substantial computational demands on downstream applications due to challenges in storage, transmission, and real-time rendering.

To alleviate these challenges of dense representations, researchers have explored the generation of these proxies, including procedural modeling, geometric simplification techniques \cite{garland1997surface,salinas2015structure,gao2022low,fastinstance2024}, and manual creation by skilled modelers.
These approximations, which trade fine geometric detail for efficiency, dramatically reduce storage and transmission requirements and enable smoother real-time rendering. 

Although textures can compensate for the loss of geometric detail, preserving the detailed appearance of the original dense reconstruction is highly challenging following geometric decimation. 
Specifically, generating realistic texture maps for simplified proxies, from an unordered collection of registered images that were used to reconstruct the original dense model, remains a daunting task.

Existing image-based texture mapping methods \cite{lempitsky2007seamless,waechter2014let,sander2003multi} often struggle when applied to proxy models. These methods frequently introduce distortions in linear features, perspective inconsistencies, and visible seams. Similarly, patch-based optimization methods~\cite{zhou2014color,bi2017patch}, designed to address photometric issues, often result in blurred textures or incomplete regions, compromising visual quality. \citet{TwinTex23} proposed a method focused on generating texture maps for architectural proxies with enhanced photometric and perspective consistency. However, this approach involves lengthy, intricate procedures and relies heavily on manual fine-tuning to achieve satisfactory results. 

In this paper, we propose a novel and versatile approach for texturing architectural proxies using a large set of unordered RGB photographs of a target building. Our method leverages differentiable rendering to generate texel-level UV maps. By utilizing extrinsic camera parameters associated with the input images, we establish precise mappings between UV map texels and image pixels. The RGB value of each texel is computed by blending the corresponding pixel values from the input images, as shown in Fig.~\ref{fig:teaser}.

Unlike existing methods that directly optimize RGB values, our approach reconstructs the UV texture by optimizing key intermediate parameters. This results in a reduced search space and enables efficient optimization through differentiable rendering. Our novel texture optimization incorporates loss functions to minimize discrepancies between rendered views and input images, while ensuring the preservation of local features and global structures.

To validate the effectiveness of our method, we conduct extensive experiments on a variety of architecturally diverse buildings. Our results demonstrate the ability of our approach to produce high-quality textured proxies that preserve critical visual features and structural details, ensuring both photometric consistency and visual plausibility.

\section{Related Work}\label{sec:rw}

\paragraph{3D reconstruction.}
A typical pipeline for automatic 3D scene reconstruction from a sequence of uncalibrated 2D images consists of two steps: Structure-from-Motion (SfM)~\cite{snavely2006photo} first establishes feature correspondences across views and estimates camera poses via incremental or global bundle adjustment, and then the subsequent multi-view stereo (MVS) techniques~\cite{furukawa2009accurate} generate dense point clouds for recovering fine shapes. 
Although this approach excels in offline and large-scale scene reconstruction, its dependency on static scenes and computational intensity limits its applicability to dynamic or real-time scenarios. Simultaneous Localization and Mapping (SLAM) addresses this gap by enabling incremental 3D reconstruction while tracking camera pose in real-time, leveraging techniques such as feature-based filtering or direct methods that minimize photometric errors~\cite{mur2015orb,engel2017direct}. In recent years, learning-based approaches~\cite {Wang_2024_CVPR,wang2025vggt,kerbl20233d} have emerged as a powerful paradigm, leveraging deep neural networks to overcome the limitations of traditional methods, introducing end-to-end frameworks that directly regress 3D geometry (such as point clouds, meshes, volumetric representations, neural representations) from images.

\paragraph{Texture mapping for 3D reconstruction.} 
Traditional approaches for mesh texturing are dominated by manual paintings or procedural models~\cite{Ebert2002}, and typically require domain-specific expertise and manual efforts. For automatic texture mapping in image-based 3D reconstruction, \textsl{view blending} methods~\cite{bernardini2001high, callieri2008masked,lee2020texturefusion} choose multiple visible views for each mesh face (or vertex) and blend these views with different weighted averaging strategies to generate texture patches. Due to geometric and camera calibration errors, textures created by these methods are often blurry and contain ghosting, especially when applied to large-scale objects such as buildings. 

In contrast, \textsl{selection-based} methods~\cite{lempitsky2007seamless,wang2018seamless,gal2010seamless, waechter2014let,fu2018texture,fu2020joint} assign each face a certain view and extract a texture patch from it, where the selection of optimal view can be formulated as a discrete labeling problem. A Markov random field is constructed to solve this multi-label problem by using different data terms and smoothness terms. Finally, they conduct seam optimization to mitigate visible seams between adjacent patches, yet they remain susceptible to introducing texture-to-texture and texture-to-geometry misalignments.

To further reduce misaligned seams, \textsl{warping-based} methods jointly rectify the camera poses and geometric errors. For example,~\cite{zhou2014color} and~\cite{bi2017patch} use local image warping and patch-based optimization to optimize camera poses while handling large geometry inaccuracies, by correctly aligning the input images. However, these methods are suitable for texturing small objects and carry a high computational cost for large-scale scenes.

To summarize, the above methods can create high-resolution texture maps for dense reconstruction, but the quality of the resultant texture mapping still greatly depends on the quality of geometric reconstruction and the accuracy of camera calibration. None of them can be naively applied to generate realistic texture maps on an abstracted proxy model, because of the significant geometric differences between the proxies and ground-truth dense models.

\paragraph{Texturing proxy geometry.} 
Proxy models can be created manually in CAD software or from procedural modeling~\cite{muller2006procedural,schwarz2015advanced}, or approximated from dense reconstruction by mesh decimation~\cite{verdie2015lod,salinas2015structure,gao2022low} and primitive-based assembly~\cite{Nan2017polyfit,fang2020connect, bauchet2020kinetic,bouzas2020structure,Guo2022}. The abstracted model has significant structural differences from the original dense mesh. However, texturing such abstracted proxy models has been less explored. A few traditional approaches~\cite{sinha2008interactive, garcia2013automatic, huang20173dlite, maier2017intrinsic3d, wang2018plane} jointly reconstruct abstracted geometry and texture maps by using primitive-based constraints. They still need to select high-quality keyframes from the input image sequence and then perform substantial optimization, so blurring and ghosting artifacts cannot be avoided. 
Recently, TwinTex~\cite{TwinTex23} generated texture maps of architectural proxy considering both perspective and photometric quality. A long sequence of sub-steps is performed to generate a texture map for each proxy polygon, involving fine-tuning a multitude of parameters and requiring long running times.

With the advent of neural fields~\cite{xu2019deep,mildenhall2021nerf,xie2022neural,mueller2022instant}, data-driven methods have been proposed to synthesize texture details for low-resolution urban geometry~\cite{Kelly2018FrankenGAN, georgiou2021projective}, or generate novel-view textures by representing textures in neural networks~\cite{oechsle2019texture,siddiqui2022texturify,metzer2022latent,baatz2022nerf}. 
Although they can generate various styles of textures, they aim for generation-oriented tasks, \textsl{i.e.,} the synthesized textures lack realism compared to the original real scenes. Moreover, high-resolution photorealistic texture maps are hard to generate for Gaussian splatting methods~\cite{kerbl20233d,huang20242d} due to the high memory requirement, which makes them less applicable in our scenario.

\section{Overview}
\label{sec:overview}

\begin{figure*}[t!]
	\centering
	\includegraphics[width=.95\linewidth]{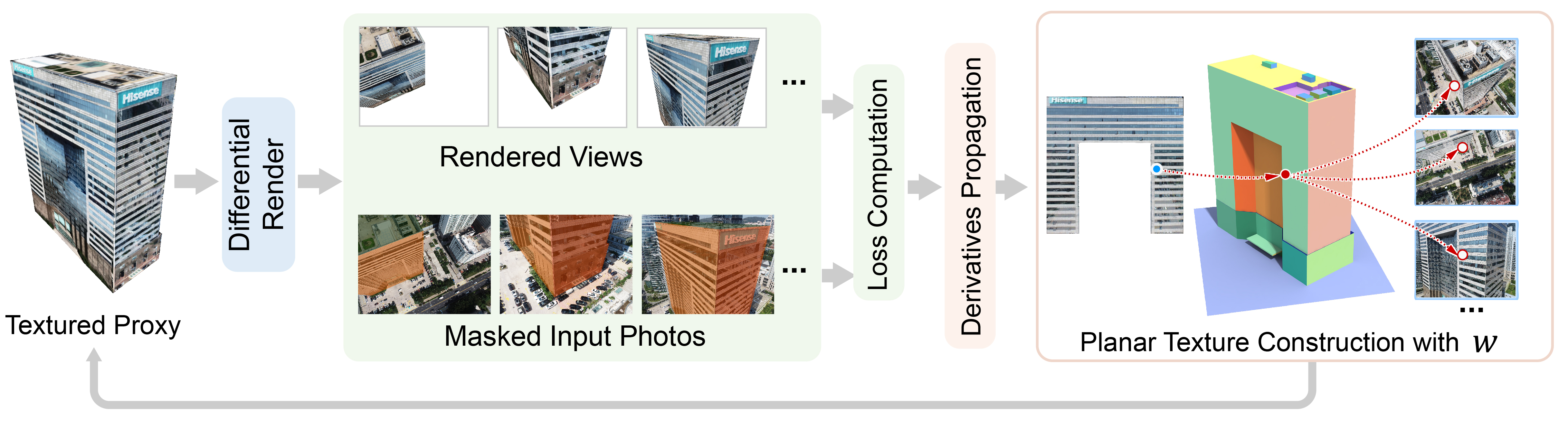}
        \vspace{-3mm}
	\caption{The differential render renders views to compute a loss, which propagates back to optimize the parameters.}
	\label{fig:pipeline}
\end{figure*}

Given an architectural proxy model represented as a set of planar \emph{proxy polygons} $\{ \mathcal{P}_i \}$, we aim at distilling a high-quality texture map $\mathcal{U}_i$, for each proxy polygon from a set of unordered, but calibrated (camera parameters are known) RGB photos $\{ \mathcal{I}_k \}$.

Each proxy polygon is assigned a 2D UV map. 
Since $\mathcal{P}_i$ is a planar polygon, rather than using a general mesh parameterization method, we define the UV map by rendering $\mathcal{P}_i$ using a virtual camera perpendicularly facing $\mathcal{P}_i$, such that the entire polygon fits in its field-of-view. 
Our goal, then, is to assign the correct color to each UV map texel.

Figures.~\ref{fig:teaser} and~\ref{fig:pipeline} show an overview of our method. In the Planar Texture Construction stage (Fig.~\ref{fig:pipeline} right), we associate each UV map texel (blue dot) with corresponding pixels in the real images (red circles).
In other words, the RGB value of each UV texel $t_o$ is obtained by blending the RGB values of its corresponding image pixels using a set of weights $\{ w_i \}$, as shown in Fig.~\ref{fig:teaser}. 
Thus, the texture reconstruction process is formulated as one of optimizing the set of blending weights for each of the texels.
In each iteration, these weights, the proxy model and the photo views together are used to reconstruct a texture map for each proxy polygon. Next, the textured proxy is fed into a differential renderer to generate a set of rendered views $\{ \mathcal{R}_k \}$ corresponding to the same camera parameters as the real input photos. 

We assess the quality of each proxy polygon texture with an objective function $L$, described below. 
The gradients of $L$ with respect to the blending weights are then computed via back-propagation, and used to update them. 
Our objective function $L$ consists of:
\begin{itemize}[leftmargin=*]
\item Render loss, $L_{Render}$, which measures the difference between the rendered views $\{ \mathcal{R}_k \}$ and their corresponding photos $\{ \mathcal{I}_k \}$. The difference is computed over the selected parts of $\mathcal{P}_i$ in each view.
\item Perspective loss, $L_{Persp}$, which prefers pixels with front-parallel viewing directions, and encourages consistent viewing directions over the entire texture map.
\item Parameter loss, $L_{Para}$, which ensures a smooth parameter field as well as a photometric consistent texture map.
\end{itemize}
Below, we describe each term of our objective function and the optimization process in more detail.

\section{Method}
\label{sec:method}

Given an unordered collection of $N_I$ photo views $\{ \mathcal{I}_k \}$ with arbitrary locations and viewing directions, our goal is to generate a set of realistic texture maps $\{ \mathcal{U}_i \}$ for an input proxy with high photometric and perspective consistency.  

\subsection{Parameters definition}
For a UV map $\mathcal{U}_i$, there is a one-to-one mapping: $f_k^i()$ from UV texel coordinates, denoted $t_{o}$, to pixel coordinates in image $\mathcal{I}_k$, denoted $p_{m}$. The mapping is given by: 
\begin{equation}
    p_{m} = f_k^i(t_{o}) = K_{k} \tilde{K}_{i}^{-1} t_{o},
\end{equation}
where the $\tilde{K}_{i}$ are the parameters of the virtual camera for the planar parameterization of polygon $P_i$ and $K_{k}$ are the parameters of the real camera that captures photo $\mathcal{I}_k$. The transformations $\tilde{K}_{i}$ and $K_{k}$ are fixed after initialization.

The color value $\mathcal{U}_i(t_{o})$ of texel $t_{o}$ can be calculated as the weighted sum of the RGB values of its corresponding pixels $\{ \mathcal{I}_k(p_{m}) \}$ from the input photo collection ($N_I$ images): 
\begin{equation}
\mathcal{U}_i(t_{o}) = \sum_{k=1}^{N_I} w_k \mathcal{I}_k(p_{m}), 
\end{equation}
where $p_{m}=f_k^i(t_{o})$, and the weights $w_k$ are in $[0,1]$.

To sum up, we optimize a normalized vector of $N_I$ weights $w_k$ for each texel in $\mathcal{U}_i$. Thus, each texture map $\mathcal{U}_i$ has a parameter field $\mathcal{W}^i_k$ for $\mathcal{I}_k$.

\subsection{Optimization function}

The pipeline of our iterative method is illustrated in Fig.~\ref{fig:pipeline}. For each proxy polygon $\mathcal{P}_i$, our method starts with an initial texture map $\mathcal{U}_i^0$ which is empty. 
We create a set of binary activation masks $\{ a_k^i \}$, each with the same size as $\mathcal{I}_k$ to indicate the selected regions of $\mathcal{I}_k$ in generating $\mathcal{U}_i$. Initially, $a_k^i$ indicate the regions of $\mathcal{P}_i$ visible in the input photo $\mathcal{I}_k$.

The weight $w_k$ of $\mathcal{I}_k$ at texel $t_{o}$ is initialized to $Q(t_o) = Q_{a} Q_{c}$. 
The quality of the viewing angle, $Q_{a}$, measures the level of frontality of photo $\mathcal{I}_k$ with respect to $\mathcal{P}_i$. 
The quality of color consistency, $Q_{c}$, measures the similarity of pixel color $\mathcal{I}_k(p_{m})$ to the dominant color among all the corresponding pixels $\{ \mathcal{I}_k(p_{m}) \}$ at $t_{o}$. 
We let $Q_{c}=Gaussian(\mathcal{I}_k(p_{m}), \mu, \sigma)$, where $\mu$ and $\sigma$ is the mean and variance of $\{ \mathcal{I}_k(p_{m}) \}$.

After each iteration, we update $\{ a_k^i \}$ by marking the selected pixel (with $w_k$ greater than a very small threshold $\tau$) of $\mathcal{I}_k$ to one, and update to zero if unselected (with $w_k \le \tau$).

\begin{figure}
	\centering
	\includegraphics[width=\linewidth]{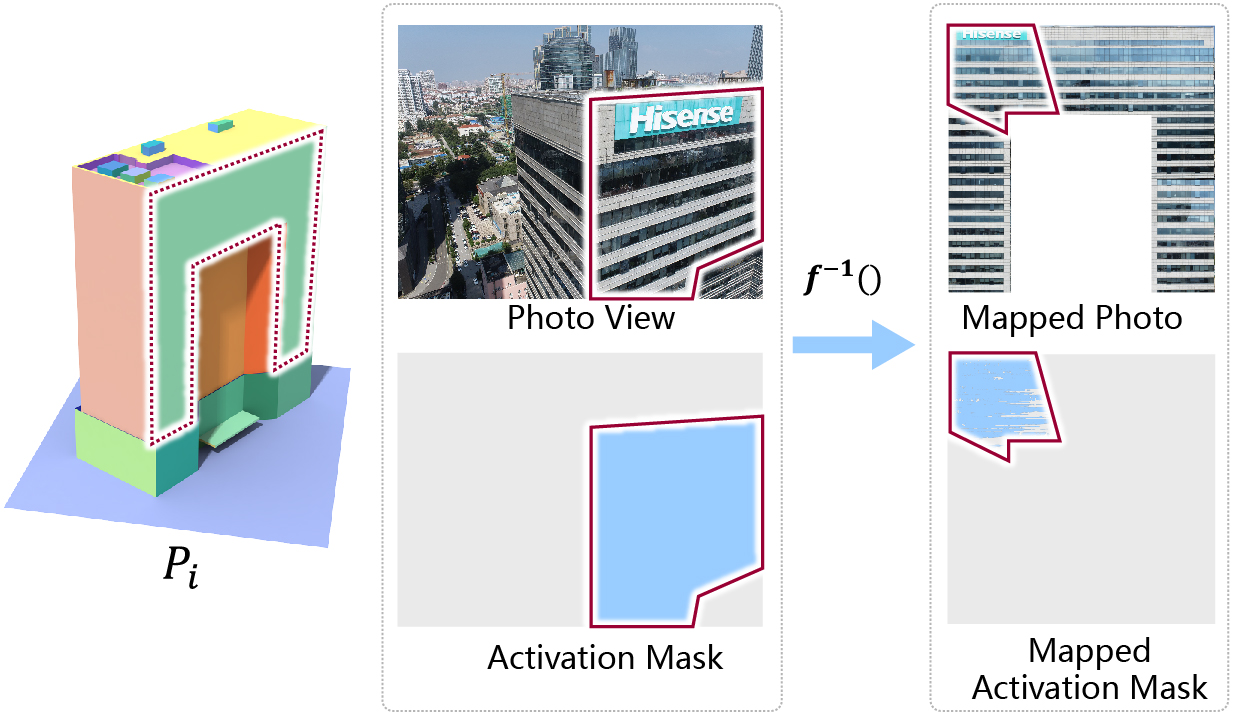}
        \vspace{-3mm}
	\caption{The photo $\mathcal{I}_k^i$ with binary activation mask $a_k^i$ (left), and the mapped photo $\widehat{\mathcal{I}}_k^i$ with mapped $\widehat{a}_k^i$ (right) against proxy polygon $\mathcal{P}_i$ of Hisense.}
	\label{fig:projection}
\end{figure}

We separately optimize the UV maps of different proxy polygons in parallel. Without loss of generality, we here describe the optimization details of reconstructing the UV map $\mathcal{U}_i$ of proxy polygon $\mathcal{P}_i$. 
At iteration $t$, given the parameters, we first construct each texel in the UV map $\mathcal{U}_{i}^{t}$ via the mapping function. 
Subsequently, all the UV texture maps and the proxy model are employed to generate a textured proxy. The textured proxy is then fed to a differentiable renderer. 
Next, we measure the quality of the reconstructed texture map $\mathcal{U}_i$ via a loss function as the combination of render loss, perspective loss and parameter loss:
\begin{equation}
L_i= \alpha L_{Render} + \beta L_{Persp} + \omega L_{Para},
\end{equation}
where $\alpha$, $\beta$ and $\omega$ are balancing coefficients for the three loss terms, with default values set to 1, 2, and 10, respectively. 
At each stage, the parameters are gradually updated to minimize $L_i$ via derivative propagation. 

\paragraph{Render Loss.}
To generate realistic texture maps ${ \mathcal{U}_i }$ for the proxy, we need to minimize the visual discrepancy between the textured proxy and the actual scene. This involves ensuring that the rendered views of the proxy match the input photographs as closely as possible.
The rendering loss $L_{Render}$ quantifies the similarity between the rendered views and the actual photos, defined as the sum of pixel-wise squared differences across all selected pixels and a regularization term:
\begin{equation}
L_{Render}= \sum_{k} ( || a_k^i  \odot Q^i \odot (\mathcal{R}_k - \mathcal{I}_k)||_2^2 + \sum_o |\mathcal{W}^i_k(t_o)| ),
\label{eq:lrender}
\end{equation}
where $\odot$ is the element-wise product, $Q^i$ is a quality matrix, matching the size of $\mathcal{I}_k$, that emphasizes high-quality pixels in the actual photo while eliminating low-quality ones. The regularization term encourages sparsity on the weights of each texel.
Each element in $Q^i$ is equal to $Q(t_o)$. The computation of $L_{Render}$ is differentiable, thus making it possible to back-propagate gradients.

\paragraph{Perspective Loss.}
The photo collections used for building reconstruction often display significant variations in viewing directions, which are vital for camera calibration. However, capturing the same content from different angles can lead to inconsistencies, resulting in ghosting and stretching issues in the reconstructed texture maps. This is particularly problematic for convex/concave facade elements and reflective surfaces such as balconies and windows.

To address this, ideally, the photos used should be taken from a fronto-parallel viewing direction. However, in practice, the input photos often exhibit significant variations in viewing angles, and photos taken from a front direction, or close to it, might not be available. 
Thus, if we consider the facade normal as a strict constraint, the selected photos may exhibit obvious missing regions.

Therefore, we introduce a guidance viewing direction considering both the consistency of viewing direction and facade normal. The direction vector for the UV map $\mathcal{U}_i$ of facade $\mathcal{P}_i$ is denoted as $\widetilde{v}_i$, which is the average vector of the inverse facade normal and the average viewing direction of visible photos: $\widetilde{v}_i=0.5*\sum v_k/|\sum v_k| - 0.5*n_i$, where $v_k$ is the viewing direction of photo $\mathcal{I}_k$.

We propose a perspective loss, $L_{Persp}$, to penalize the mapped pixels whose camera directions differ significantly from the guidance viewing direction: 
\begin{align}
\label{eq:lpersp}
&L_{Persp}= \sum_{k} || \mathcal{W}_k^i
\odot \mathcal{V}_k^i \odot \mathcal{V}_k^i||_1,\\
&\mathcal{V}_k^i = \widehat{a}_k^i \odot (\widehat{\mathcal{V}}_k - \widetilde{\mathcal{V}}_i).
\end{align}
Here, $\widehat{a}_k^i$ is the mapped activation mask of $a_k^i$ on $\mathcal{P}_i$ (see Fig.~\ref{fig:projection}), $\widehat{\mathcal{V}}_k$ indicates the viewing direction field of photo $\mathcal{I}_k$, matching the size of $\mathcal{U}_i$. Each element in $\widehat{\mathcal{V}}_k$ corresponds to the inverse viewing direction of the corresponding pixel in $\mathcal{I}_k$ relative to the camera position. Each element in $\widetilde{\mathcal{V}}_i$ is equal to $\widetilde{v}_i$.

\begin{figure*}
	\centering
	\includegraphics[width=.95\linewidth]{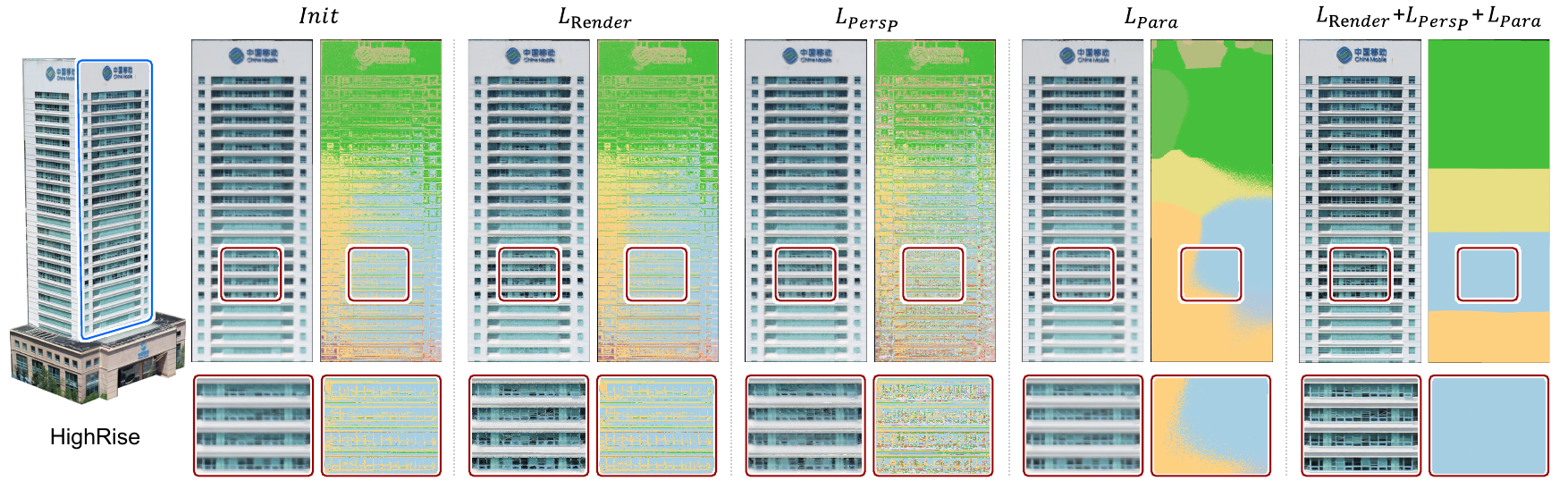}
    \vspace{-3mm}
	\caption{{The initial facade texture maps, and the initial texture maps optimized with $L_{Render}$, $L_{Persp}$, $L_{Para}$, $L_{Render} + L_{Persp} + L_{Para}$ on HighRise example are visualized from left to right. For each facade texture, a texel source map visualizing the photo source (with dominant $\mathcal{W}$ value) of all texels is placed on its right. Each input photo is represented by a unique, randomly assigned color.}}
	\label{fig:highrise-ablation}
\end{figure*}

\begin{figure*}[tbh!]
	\centering
	\includegraphics[width=.95\linewidth]{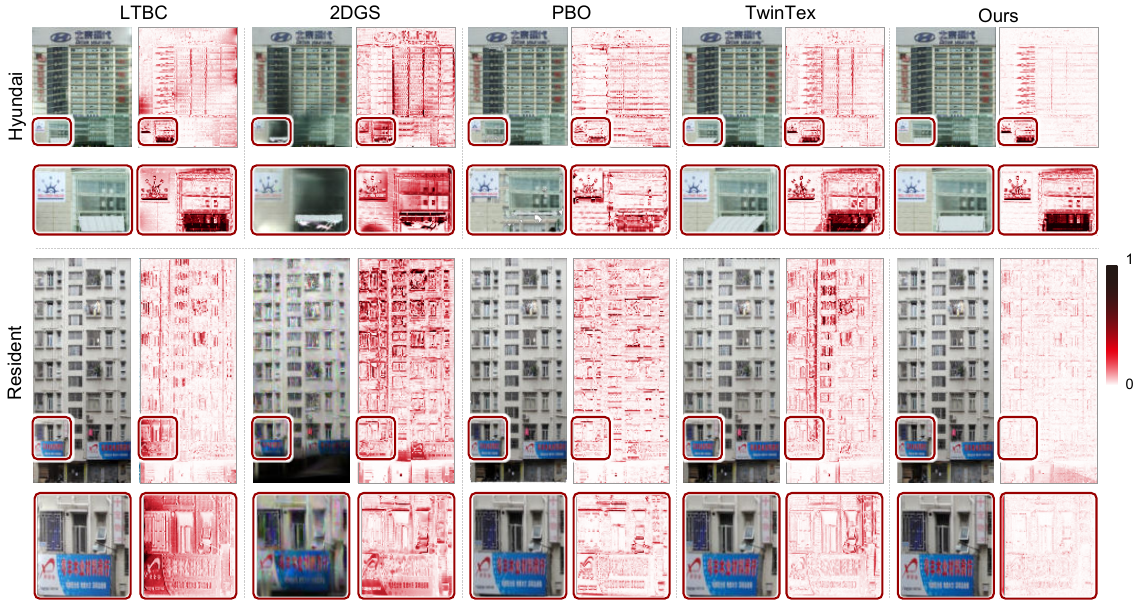}
	\caption{{Comparison with LTBC, 2DGS, PBO, TwinTex and ours on facades of two virtual buildings. The texel-wise differences compared with the GT maps are visualized on the right. The colors denote error values from low (white) to high (dark red).}}
	\label{fig:comparison-virtural}
\end{figure*}

\begin{table*}
    \caption{{Quantitative comparison of the texturing performance on the virtual dataset. Numbers denote the quality values of generated texture maps.}} 
    \label{tab:quality-results-virtual}
    \begin{center}
		\begin{tabular}{c c c c c c c}
			\midrule[1pt]
			Scene & Method & Error (90\%) $\downarrow$ & Error (95\%) $\downarrow$  & $SSIM \uparrow$ & $LPIPS \downarrow$    & Time (min) \\
			\midrule
		\multirow{5}{*}{Hyundai}   
			& LTBC  & 0.311  & 0.439 &  0.831  &  \textbf{0.124}   & 0.4         \\       
			&  2DGS  &  0.449 & 0.604   &  0.689   &    0.323 & 20.4           \\    
                & PBO  & 0.292 & 0.425  & 0.761    &   0.199  &   32.7       \\ 
                & TwinTex  & 0.276 & 0.388 &  0.779   &  0.178    &    1.3       \\ 
                &  Ours  & \textbf{0.128} & \textbf{0.277}   &  \textbf{0.881} & 0.130   & 1.2         \\
   			\midrule
		\multirow{5}{*}{Resident}   
			& LTBC  & 0.198  &  0.284 &  0.807   &  \textbf{0.067}   & 0.4         \\       
			&  2DGS   & 0.588 & 0.717  & 0.350   &  0.535   & 19.4           \\              
                & PBO  & 0.227 & 0.367 &  0.675   &  0.183    &  13.9       \\ 
                & TwinTex  & 0.281 & 0.409 & 0.669  &  0.173    &   10.9       \\ 
                &  Ours  & \textbf{0.066} & \textbf{0.092} &  \textbf{0.920}  & 0.074   & 1.2         \\ 
                \midrule[1pt]
		\end{tabular}
    \end{center}
\end{table*}%

\paragraph{Parameter Loss.}
Optimizing the parameters of neighboring texels independently ignores the coherence of contents and introduces noise. Ensuring that neighboring texels are derived from similar sets of photos can maintain the consistency of content. 
However, simply insisting that neighboring texels should originate from the same photo can lead to noticeable seams. The Graph-cut algorithm~\cite{kwatra2003graphcut} aims to find a cut in the overlapping region of two images for seamless color transition. We propose a loss, $L_{Para}$, to encourage parameter smoothness while penalizing neighboring texels exhibiting large color and gradient variations in the overlapping region: 
\begin{align}
& L_{Para} = \sum_{o} \sum_{q \in N(o)}
C(t_o, t_q) |\widehat{a}_k^i(t_o) \mathcal{W}_k^i(t_o) - \widehat{a}_k^i(t_q) \mathcal{W}_k^i(t_q)|, \\
& C(t_o, t_q)= m_k^i(t_o) D(t_o, t_q) + \lambda_s, \\
& D(t_o, t_q) = S(|| (\mathcal{U}_i(t_o) - \mathcal{U}_i(t_q)) ||_2 ) (1 + G(t_o)),
\end{align}
where $N(o)$ is the up and left neighbors of texel $t_{o}$.  $\widehat{a}_k^i(t_o)$ is the binary value of $\widehat{a}_k^i$ at $t_o$, $\widehat{\mathcal{I}}_k^i$ is the mapped $\mathcal{I}_k^i$ on $\mathcal{P}_i$, $m_k^i$ is the mask denoting the overlap of $\widehat{\mathcal{I}}_k^i$ to all the other mapped photos (one for overlap while zero otherwise), ${\mathcal{W}}_k^i(t_q)$ is the value of ${\mathcal{W}}_k^i$ at $t_q$, $S(x)$ is the sigmoid activation function, $G(t_o)$ is the mean of gradient magnitude of all the mapped photos at texel $t_o$. 
We set $\lambda_s=0.5$ in our experiments to balance smoothness and disparity among parameters of neighboring texels.
{$L_{Para}$ is computed over all mapped images for the texel (summing over $k$). An element in $m^i_k$ is set to 1 if any other mapped image overlaps with $Î^i_k$.}

\subsection{Data Quality Optimization}

\paragraph{Data Elimination.}
In practice, the input photos often exhibit significant variations in photometric and perspective quality. 
Simply taking the average of all the corresponding pixels of the target texel introduces severe ghosting and blurring issues. 
We aim to select only a few semantically coherent photos, thereby letting only a few \textit{key pixels} contribute to the final color value of one texel.
To achieve the above goals, we progressively optimize the input data by: i) excluding inclined and perspective inconsistent photos, and $\mathcal{I}_k$ from $\{ \mathcal{I}_k \}$ if the portion of zero value weight in $\mathcal{W}^i_k$ (regarding the mask) exceeds a threshold $\tau_w$, ii) eliminating the selected pixel for texel $t_{o}$ if its color is distinct from the average color of all the selected pixels observing $t_{o}$. More specifically, for texel $t_o$, we compute the mean and standard deviation of its mapped pixel colors, removing any pixels lying outside three standard deviations. We indicate the eliminated pixels in the activation mask as zero.

Improper photo or pixel removal can cause irreversible color loss, as the optimization process cannot infer missing regions. We perform a stress test to validate the effectiveness of our \textit{Data Elimination} by texturing the proxy model given various random subsets of input photos. Please refer to Section D of the supplementary document for more details.

\paragraph{Coarse-to-Fine Resolution.}
Our texture generation process follows a coarse-to-fine strategy to minimize computational costs while progressively producing higher-resolution texture maps. Initially, the maximum resolution of the target UV map is set to 256. Typically, neighboring texels share similar parameters. After convergence, we double the texture resolution by upscaling the parameter field using bilinear interpolation. {This upscaling process is repeated until the target resolution is reached, i.e., 2048}.

\paragraph{Brightness Adjustment.}
Since each texture map is composed of pixels from various viewing directions, the brightness may be inconsistent across neighboring texels. We perform a histogram matching 
operation on all the photos to improve the illumination consistency of $\mathcal{U}_i$ in HSV color space.

\section{Results}
\label{sec:results}

\paragraph{Implementation.}
Our method is implemented with Nvdiffrast~\cite{laine2020modular}. All the presented experimental results are obtained on a desktop computer equipped with an Intel i9-12900k processor with 3.0 GHz, 128 GB RAM and an NVIDIA RTX 3090Ti GPU.
We begin with a filtering step that eliminates all the invisible photos, blurry photos, and photos with extremely inclined angles from the inputs.
We use the Adam optimizer~\cite{kingma2014adam} with default parameters of $\alpha = 0.005$, $\beta_1 = 0.9$, and $\beta_2 = 0.99$, with a constant learning rate throughout training. Additionally, we set $\tau_w = 0.95$ in our experiments.

\paragraph{Dataset}
We evaluate the quality of texture maps with synthetic models and real-world buildings (including $14$ outdoor buildings, two outdoor scenes and an indoor scene) in this section and supplemental material. 
We first construct a synthetic dataset containing five virtual models (Hall, Mall, Resident, Archive and Hyundai~\cite{UrbanScene3D}). All the models are created by an in-field modeler. For each model, we rendered $200$ views to cover the entire building, all with precise camera parameters without any feature distortions. 
The real-world scenes come from public sources or are captured by a single-camera drone~\cite{DroneScan20,TwinTex23}.
We use the commercial software RealityCapture\footnote{https://www.capturingreality.com/} (RC) to reconstruct the 3D high-precision models from the captured images.
All the abstracted versions are either the 2.5D models for path planning or created by in-field modelers.
All of the dense reconstructions and proxy models are utilized to demonstrate the performance of our algorithm. 
Please refer to the supplementary material for more details and self-evaluation experiments.

\subsection{Ablation Study}

We utilize a real-world example, called HighRise, to validate the effectiveness of each loss term. The texturing results are visualized in Fig.~\ref{fig:highrise-ablation}. 
We can observe that the initialized texture map is blurry and has ghosting issues. One reason is the existence of common errors in the registered camera parameters. 
The optimized texture map becomes sharper with only $L_{Render}$, which effectively eliminates photo pixels with large camera errors and photometric disparity. Even so, the results still suffer from blurring issues, as shown in the second column in Fig.~\ref{fig:highrise-ablation}. 
Another reason is that, in practice, the captured photos often exhibit biased viewing directions.
Most photos are typically taken from a top-down perspective. Large variations in viewing directions can exist as well. Such variation will bring local blurring issues to regions with convex/concave facade elements. 
We can alleviate such a problem with only $L_{Persp}$, \eg the texture maps on the third column in Fig.~\ref{fig:highrise-ablation} reveal higher sharpness. However, the result reveals noisy texels and a lack of content continuity. 
We then optimize $\mathcal{W}$ with only $L_{para}$, encouraging a smooth parameter field. The texture map on the fourth column in Fig.~\ref{fig:highrise-ablation} reveals much higher sharpness compared to the first column. However, this still can not address the ghosting issues (see the bottom right region).
Our final method, involving all the key components, gives a texture map with high photometric and high perspective quality.
A stress test in the supplementary document (Section D) demonstrates that our method effectively generates high-quality texture maps even under extremely limited resources.

\begin{table*}
    \caption{Quantitative comparison on the performance of selected photos and texturing results on real datasets. Numbers in the brackets denote the quality value of the zoomed-in views of each example from top to bottom.}
    \label{tab:quality-results-real}
    \begin{center}
    \resizebox{.8\textwidth}{!}{
	\begin{tabular}{c c c c c c c }
		  \midrule[1pt]
		  \multirow{2}{*}{Scene} & \multirow{2}{*}{Method}  & \multicolumn{2}{c}{Perspective Quality}   & \multicolumn{2}{c}{Overall Quality} & {Time}\\
            \cmidrule(lr){3-4} \cmidrule(lr){5-6}
            & & $Q_{front} \uparrow$  & $Q_{vc} \uparrow$  & $SSIM \uparrow$ & $LPIPS \downarrow$    & (min) \\
            \cmidrule(lr){1-2} \cmidrule(lr){3-4} \cmidrule(lr){5-6} \cmidrule(lr){7-7}
            \multirow{5}{*}{{Center}}   & {RC}   & - & -  &  \{0.36, 0.40, 0.60, 0.54\}    &   \{\textbf{0.49}, 0.43, 0.35, 0.17\}      & {287.0}       \\
            & {LTBC} & \{0.71, 0.77, 0.71, 0.69\}& \{0.59, 0.39, 0.59  0.48\} &  \{0.28, 0.34, 0.27, 0.50\}  &      \{0.58, 0.52, 0.32, 0.18\} & {1.1}        \\ 
		  & {PBO}  & \{0.52, 0.61, 0.52,0.72\} & \{0.33, 0.62, 0.33, 0.51\} &  \{0.17, 0.35, 0.62, 0.55\}   &    \{0.67, 0.62, 0.36, 0.25\}    & {124.9}         \\
		  &  {TwinTex}  & \{0.89, \textbf{0.98}, 0.89, \textbf{0.98}\}& \{0.78, 0.53, 0.78, 0.77\}  &   \{0.12, 0.40, \textbf{0.64}, 0.65\}  &   \{0.69, 0.49, 0.40, 0.15\}    & {27.2} \\
            &  {Ours}  & \{\textbf{0.91}, 0.93, \textbf{0.91}, 0.92\}&  \{\textbf{0.88}, \textbf{0.89}, \textbf{0.88}, \textbf{0.81}\}  &  \{\textbf{0.52}, \textbf{0.54}, \textbf{0.64}, \textbf{0.66}\}  &   \{0.52, \textbf{0.34}, \textbf{0.27}, \textbf{0.11}\}    & {11.4}     \\  
            \cmidrule(lr){1-2} \cmidrule(lr){3-4} \cmidrule(lr){5-6} \cmidrule(lr){7-7}
            \multirow{5}{*}{{Library}}   & {RC}   & - & - &  \{ 0.56, 0.65, 0.06, 0.17\}    &   \{\textbf{0.44}, 0.39, 0.68, 0.60\}      & {383.6}       \\
            & {LTBC} & \{0.78, 0.86, 0.67, 0.82\} & \{0.66, 0.39, 0.25, 0.27\}  &  \{ 0.54, 0.47, 0.14, 0.65\}  &      \{0.45, 0.53, \textbf{0.27}, 0.37\} & {1.0}        \\ 
		  & {PBO} & \{0.94, 0.85, 0.98, 0.89\} & \{0.93, 0.52, 0.67, 0.27\}  &  \{ \textbf{0.63}, 0.72, \textbf{0.19}, 0.69\}   &    \{ 0.54, 0.46, 0.49, 0.40\}    & 106.4        \\
		  &  {TwinTex}  & \{\textbf{0.99}, \textbf{0.96},\textbf{ 0.99}, 0.83\}& \{0.97, 0.57, \textbf{0.71}, 0.80\}    &   \{0.59, 0.65, 0.16, 0.70\}  &   \{ 0.46, 0.30, 0.33, 0.39\}    & {35.8} \\
            &  {Ours}  & \{\textbf{0.99}, 0.93, \textbf{0.99}, \textbf{0.90}\} & \{\textbf{0.98}, \textbf{0.62}, 0.67,\textbf{ 0.90}\}   &   \{ 0.55, \textbf{0.73}, 0.18, \textbf{0.75}\}  &   \{ 0.47, \textbf{0.22}, 0.31, \textbf{0.29}\}    & {11.7}     \\ 
            \cmidrule(lr){1-2} \cmidrule(lr){3-4} \cmidrule(lr){5-6} \cmidrule(lr){7-7}
            \multirow{5}{*}{{Hisense}}   & {RC}   & - & -  &  \{0.52, 0.46, 0.44, 0.45\}    &      \{0.32, 0.40, 0.49, \textbf{0.42}\}\}    & {792.3}       \\
            &\small{LTBC} & \{0.70, 0.70, 0.51, 0.74\}& \{0.70, 0.70, 0.38, 0.73\}  &  \{0.43, 0.38, 0.35, 0.35\}  &      \{0.40, 0.44, 0.51, 0.45\} & {1.7}        \\
		  & {PBO} & \{0.79, 0.79, 0.67, 0.61\} &  \{0.81, 0.81, 0.38, 0.63\} &  \{0.48, 0.48, \textbf{0.46}, \textbf{0.46}\}   &     \{0.40, 0.36, 0.51, 0.48\}    & {144.6}         \\    
		  &  {TwinTex}  & \{0.86, 0.86, 0.84, 0.54\}& \{0.85, 0.85, 0.31, 0.54\} &   \{0.56, \textbf{0.49}, 0.42, 0.41\}  &   \{\textbf{0.31}, 0.34, 0.51, 0.46\}    & {33.7} \\
            &  {Ours}  & \{\textbf{0.92}, \textbf{0.92}, \textbf{0.93}, \textbf{0.96}\}& \{\textbf{0.92}, \textbf{0.92}, \textbf{0.90}, \textbf{0.96}\}  &   \{\textbf{0.57}, 0.48, 0.40, 0.42\}   &     \{0.32, \textbf{0.33}, \textbf{0.48}, 0.46\}   & {14.6} \\
            \cmidrule(lr){1-2} \cmidrule(lr){3-4} \cmidrule(lr){5-6} \cmidrule(lr){7-7}
            \multirow{5}{*}{Highrise}   & {RC}   & - & - &  \{0.52, 0.31, 0.46, 0.36\}    &   \{0.29, 0.45, 0.49, 0.47\}      & {123.0}       \\
            & {LTBC} & \{0.66, 0.66, 0.79, 0.67\}& \{0.36, 0.36, 0.56, 0.46\}  &  \{0.50, 0.32, 0.50, 0.36\}  &      \{0.27, 0.46, \textbf{0.43}, \textbf{0.46}\} & {0.8}        \\   
		  & {PBO}  & \{0.77, 0.77, 0.75, 0.92\} & \{0.79, 0.79, 0.69, 0.93\} &  \{0.53, \textbf{0.48}, \textbf{0.53}, \textbf{0.40}\}   &    \{0.39, 0.43, 0.52, 0.60\}    & {64.7}         \\ 
		  &  {TwinTex}   &\{\textbf{0.95}, \textbf{0.95}, \textbf{0.92}, 0.98\} & \{0.92, 0.92,0.66, 0.97\} &   \{0.55, 0.35, 0.52, 0.35\}  &   \{0.28, 0.47, 0.45, 0.50\}    & {5.3} \\
            &  {Ours}  & \{\textbf{0.95}, \textbf{0.95}, 0.86, \textbf{0.99}\} & \{\textbf{0.94}, \textbf{0.94}, \textbf{0.70}, \textbf{0.98}\} &   \{\textbf{0.58}, 0.45, 0.52, 0.35\}  &   \{\textbf{0.25}, \textbf{0.34}, 0.47, 0.49\}    & {4.0}             \\
            \cmidrule(lr){1-2} \cmidrule(lr){3-4} \cmidrule(lr){5-6} \cmidrule(lr){7-7}
  		\multirow{5}{*}{{PolyTech}}   & {RC}  & - & -  &  \{0.33, 0.22, 0.58, 0.65\}    &   \{0.44, 0.19, 0.24, 0.18\}      & {287.0}       \\
		  & {LTBC}  &\{0.55, 0.71, 0.48, 0.88\} &\{0.53, 0.47, 0.33, \textbf{0.87}\}  &  \{0.34, 0.22, 0.56, 0.68\}  &      \{0.39, 0.23, 0.21, 0.31\} & {4.4}        \\ 
            & {PBO}  & \{0.68, 0.28, 0.74, \textbf{0.94}\}  &  \{0.55, 0.32, 0.60, 0.79\}  &   \{0.36, 0.22, 0.56, 0.53\}   &    \{0.36, 0.48, 0.55, 0.39\}    & {180.3}         \\
		  &  {TwinTex}  &\{0.54, \textbf{0.95}, 0.69, 0.85\}  &\{0.49, 0.56, 0.40, 0.77\}   &   \{\textbf{0.42}, 0.25, 0.58, 0.66\}  &   \{0.38, 0.33, 0.23, 0.20\}    & {74.2} \\
    	&  {Ours}  &\{\textbf{0.90}, 0.92, \textbf{0.89}, 0.91\}  & \{\textbf{0.67}, \textbf{0.66}, \textbf{0.68}, 0.74\} &   \{0.38, \textbf{0.26}, \textbf{0.65}, \textbf{0.70}\}  &   \{\textbf{0.28}, \textbf{0.18}, \textbf{0.16}, \textbf{0.15}\}    & {14.1}     \\
        \midrule[1pt]
        \end{tabular}
       }
	\end{center}
\end{table*}%

\subsection{Comparisons on synthetic Scenes}

In this section, we compare our method against multiple methods on both synthetic and real architectural models.
Due to the difference in the heuristics among these methods, we perform resampling~\cite{wang2021real} to guarantee their inputs and results share the same resolution.
Two visual similarity metrics, structural similarity index measure (SSIM) and learned perceptual image patch similarity (LPIPS)~\cite{kastryulin2022piq}, are adopted for quantitative evaluation.

\paragraph{Qualitative comparisons.}
Firstly, we compare our method against let there be color (LTBC) \cite{waechter2014let}, 2D Gaussian Splatting (2DGS) \cite{huang20242d}, patch-based optimization (PBO)~\cite{bi2017patch} and TwinTex~\cite{TwinTex23} on synthetic models with precise camera parameters.

For each planar polygon of a virtual model, we regard its UV texture map as the ground truth map. We use the evaluation scheme to compare the texture maps or screenshots to the corresponding ground truth maps. Qualitative results are visualized in Fig.~\ref{fig:comparison-virtural}. Error maps of resultant maps are provided as well. 
All the methods achieve feasible results in most cases. 
However, both LTBC and 2DGS introduced texels with abnormal brightness.
All results in Fig.~\ref{fig:comparison-virtural} are generated by using the same input set without brightness adjustment. LTBC applies its own adjustment, explaining its abnormal brightness. Our method and 2DGS do not use such an adjustment.
The comparison shows that given the same set of views, 2DGS can generate a scene with satisfying quality in viewing points similar to the input. However, as we move close to the facade at an unseen angle, the results of 2DGS can show cotton-like or needle-like structures. 
PBO is less stable and sometimes produces empty regions, as shown in the Resident example. Overall, both our method and TwinTex yield plausible results in these cases. However, our approach produces texture maps with significantly fewer errors compared to LTBC, 2DGS, PBO, and TwinTex.

\paragraph{Quantitative comparisons.}
Quantitative results are listed in Table~\ref{tab:quality-results-virtual}.
First, we compute \textit{Error} to assess the quality of the reconstructed texture map by measuring its similarity to the ground truth. Specifically, for each texel, the error is defined as the distance to its corresponding ground-truth color. We then analyze the distribution of these error values and report a threshold $x$ as the error statistic, where $x$ indicates that 95\% of texels have a distance smaller than this value.
We can see that our Errors are significantly smaller than those of other methods.
Moreover, our UV texture maps deliver better SSIM and LPIPS scores~\cite{zhang2018unreasonable}. Please refer to the supplementary material for more results.
Since our methods and LTBC share different representations of the reconstructed results and tasks compared to 2DGS, the quantitative results may not be fair for either side. Therefore, in subsequent experiments, we only perform comparisons to texturing methods (\ie excluding 2DGS) that share the same representation to ensure fairness.

\begin{figure*}
	\centering
	\includegraphics[width=\linewidth]{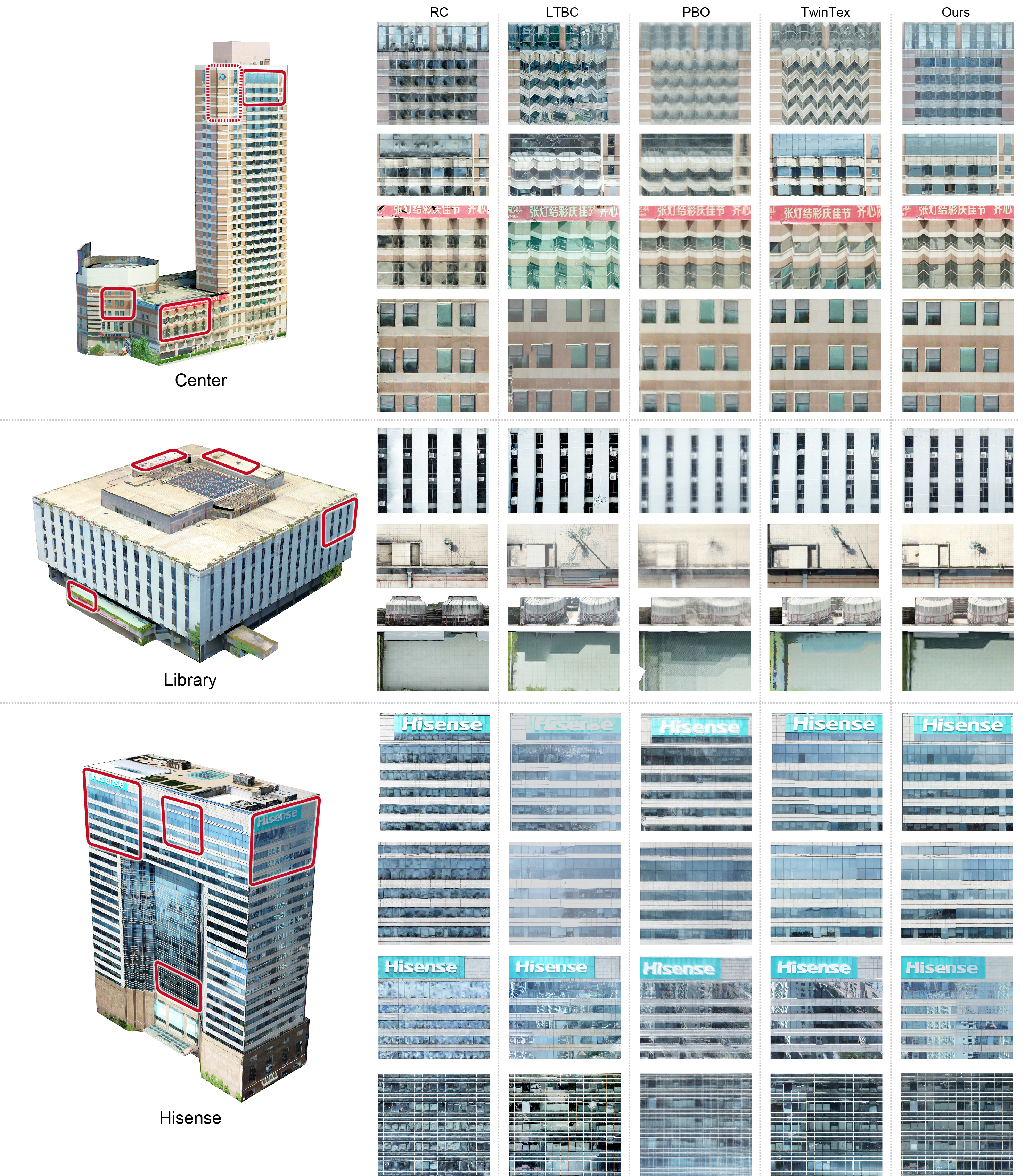}
	\caption{Comparison of texturing results with RC, LTBC, PBO, TwinTex, and ours on real-world buildings.}
	\label{fig:comparison2}
\end{figure*}

\begin{figure*}
	\centering
	\includegraphics[width=\linewidth]{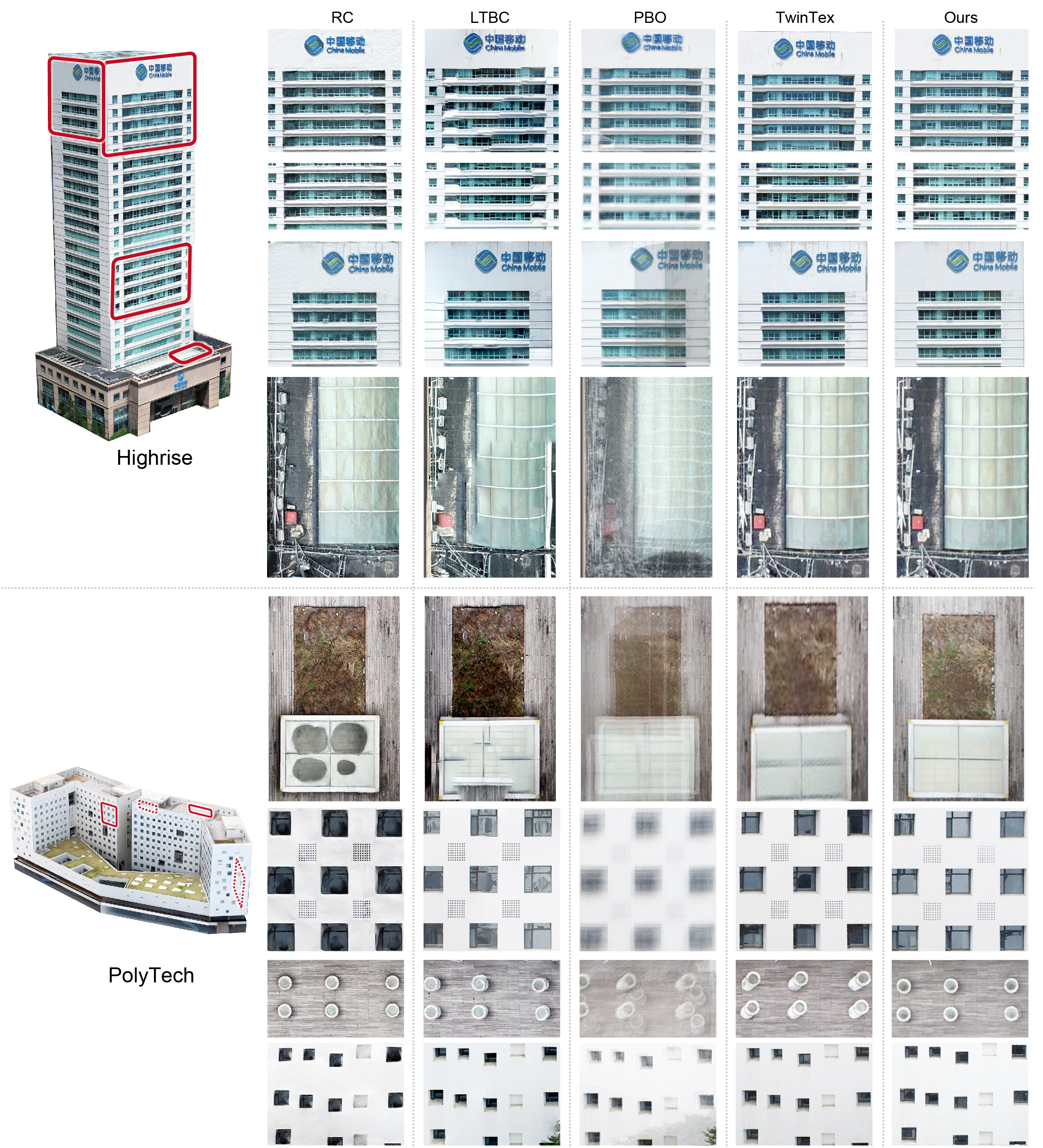}
	\caption{Comparison of texturing results with RC, LTBC, PBO, TwinTex, and ours on real-world buildings.}
	\label{fig:comparison}
\end{figure*}

\begin{figure*}
	\centering
	\includegraphics[width=.96\linewidth]{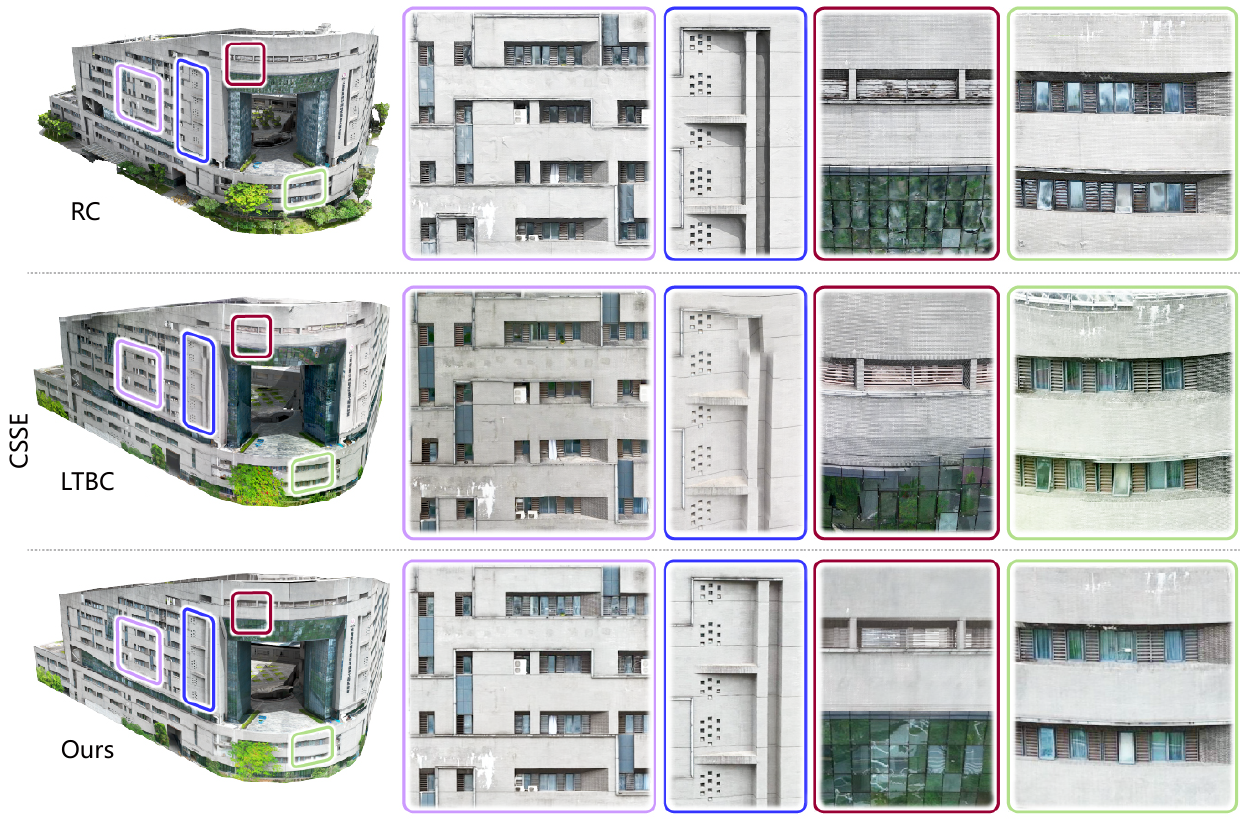} 
    \caption{Comparison of texturing results with RC, LTBC and ours on a challenging CSSE building. The details are shown in the zoomed-in insets.}
    \label{fig:challenging-cases}
\end{figure*}

\begin{figure*}
	\centering
	\includegraphics[width=.96\linewidth]{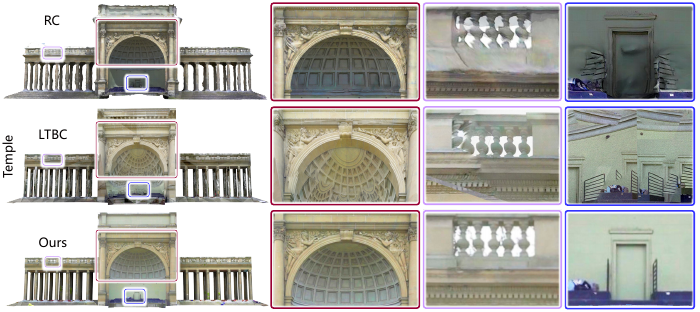}
    \caption{{Comparison of texturing results on the challenging Temple example using RC, LTBC, and our method.}}
	\label{fig:comparison-extra}
\end{figure*}

\begin{figure*}
	\centering
    \includegraphics[width=\linewidth]{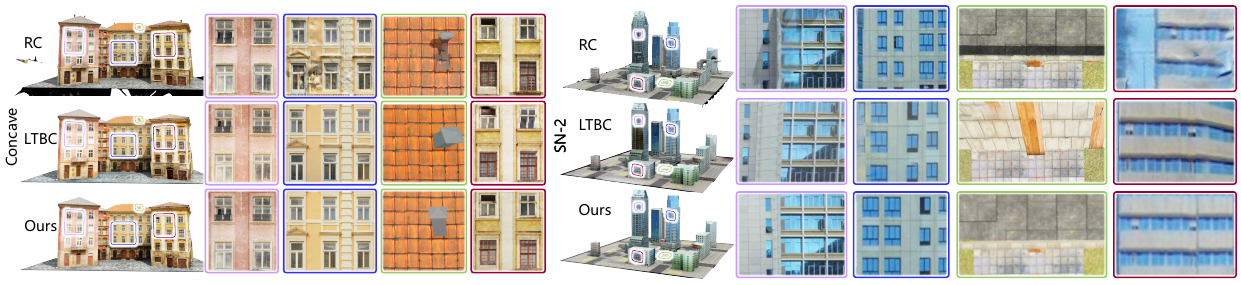}
	\caption{{Comparison of texturing results on two large scenes using RC, LTBC, and our method.}}
	\label{fig:comparison-extra2}
\end{figure*}

\subsection{Comparisons on Real Scenes}

\paragraph{Qualitative comparisons.}
For the real scenes, we compare our method against LTBC, patch-based optimization (PBO)~\cite{bi2017patch} and TwinTex~\cite{TwinTex23}. 
We adopt the evaluation scheme in ~\citet{Waechter2017} and ~\citet{TwinTex23} to compare the rendered image with the corresponding real image following a specific view from the input cameras. We select the ground truth photos for evaluation from the input excluding the photos for texturing.

Qualitative results are visualized in Fig.~\ref{fig:comparison2} and Fig.~\ref{fig:comparison}. The dense models reconstructed with RC are shown as references on the facade structure.
First, without the constraints on perspective consistency, LTBC failed in selecting photos with both consistent and front-parallel viewing directions, creating obvious seams, perspective inconsistency and “stretching" alike artifacts in the generated textures. 
The perspective inconsistency can be observed in most of the examples, such as the Center (first and second rows, second column) in Fig.~\ref{fig:comparison2}, the Higrise example (first row, second column) in Fig.~\ref{fig:comparison}. This resulted from merging photos with large viewing angle variance. 
TwinTex and our method consider both photometric and perspective consistency among selected photos. However, unlike TwinTex, our strategies can select perspective-consistent and front-parallel pixels as much as possible, which yields smoother perspective variation and better frontality in most of the examples.
Early-stage decisions that overlook final objectives can create challenges for subsequent steps. TwinTex, for instance, emphasizes minimizing the number of source photos, which may introduce boundary misalignments. In contrast, our method jointly optimizes core goals. For example, to generate the facade texture in Fig.~\ref{fig:highrise-ablation}, TwinTex uses only 3 photos, while our approach leverages more, resulting in better perspective alignment and photometric quality across the entire facade.
The “stretching" problem can be observed in the inset of the Center (first and third rows, second and fourth columns) building and the Library (six row, second and fourth columns) in Fig.~\ref{fig:comparison2}, the PolyTech (seven row, fourth column) in Fig.~\ref{fig:comparison}, which was caused by using photos with inclined viewing angles.  
We introduced $L_{Para}$ to consider the color distance of source pixels between neighboring texels. This helps generate texels with a higher level of photometric and content consistency compared to TwinTex. The significant improvement can be observed in the Center example in Fig.~\ref{fig:comparison2}. 
PBO does not design a photo selection mechanism, is sensitive to the inputs and often generates empty regions. For fair comparisons, we only use a small number of high-quality photos selected by TwinTex and ours as its input. However, PBO still generates results with blurring or ghosting issues.

\paragraph{Quantitative comparisons.}
Quantitative results on selected photos and texture maps are listed in Table~\ref{tab:quality-results-real}. 

First, we employ SSIM and LPIPS to evaluate the quality of the final texture maps. Since SSIM is by nature less sensitive to blurring issues and possibly gives higher scores to images with such artifacts~\cite{zhang2018unreasonable}, the texture with the regional blurring problem could have higher scores. This phenomenon can be observed in some views of the Highrise, Hisense and Library examples. Meanwhile, LPIPS is designed to match human perception and yield better scores for images with a higher level of coherence. In most cases, our UV texture maps can deliver better SSIM and LPIPS scores.

Next, we use the quality metrics in~\citet{TwinTex23} to measure the perspective quality of a texture map at the texel level:
i) viewing angle frontality $Q_{front}$, and ii) viewing angle consistency $Q_{vc}$.
All four quality values are normalized to $[0, 1]$. The larger value indicates higher quality. Please refer to the supplemental for more details.
Overall, the statistics indicate that our method has the best perspective consistency, while our method and TwinTex achieve the best frontality. This is because the photos are captured by a perspective camera; there are still inclined contents for pixels away from the photo center, even if the photo is taken from a front-parallel angle (especially for large facades). Hence, unlike TwinTex, which selects very few single photos to cover a large plane, we encourage texel-level texture generation, which yields better perspective quality over the entire facade.

Additionally, TwinTex is usually more than two times slower than our method. And PBO’s computational time is often more than three times that of TwinTex, eight times that of our method. Among all the methods, LTBC requires the least running time.

\subsection{Challenging Cases}

In this subsection, we experiment with more challenging examples and various building styles.
Firstly, we experiment on a challenging CSSE example that contains curved surfaces and reflective glass, as shown in Fig.~\ref{fig:challenging-cases}. 
Next, in Fig.~\ref{fig:comparison-extra}, we experiment on a temple example with thin pillars and complex structures from public sources~\cite{DroneScan20,knapitsch2017tanks}.
Finally, we experiment on two complicated scenes with multiple buildings~\cite{DroneScan20}, in Fig.~\ref{fig:comparison-extra2}.
The visual results further suggest that the textures generated by our approach exhibit high perspective consistency and harmonious illumination.

\section{Conclusion and Future Work}

In this work, we introduce an automated technique for creating realistic texture maps for architectural proxy models using unordered RGB photographs. The core of our approach is to parameterize the texturing process by establishing a correspondence between texels on a UV map and pixels in the input photograph. The color value of each texel is derived as the weighted sum of these associated pixels. We employ differentiable tracing for individual texels to refine blending parameters, ensuring photometric and perspective consistency over a coherent texture map.

\begin{figure}
	\centering
	\includegraphics[width=\linewidth]{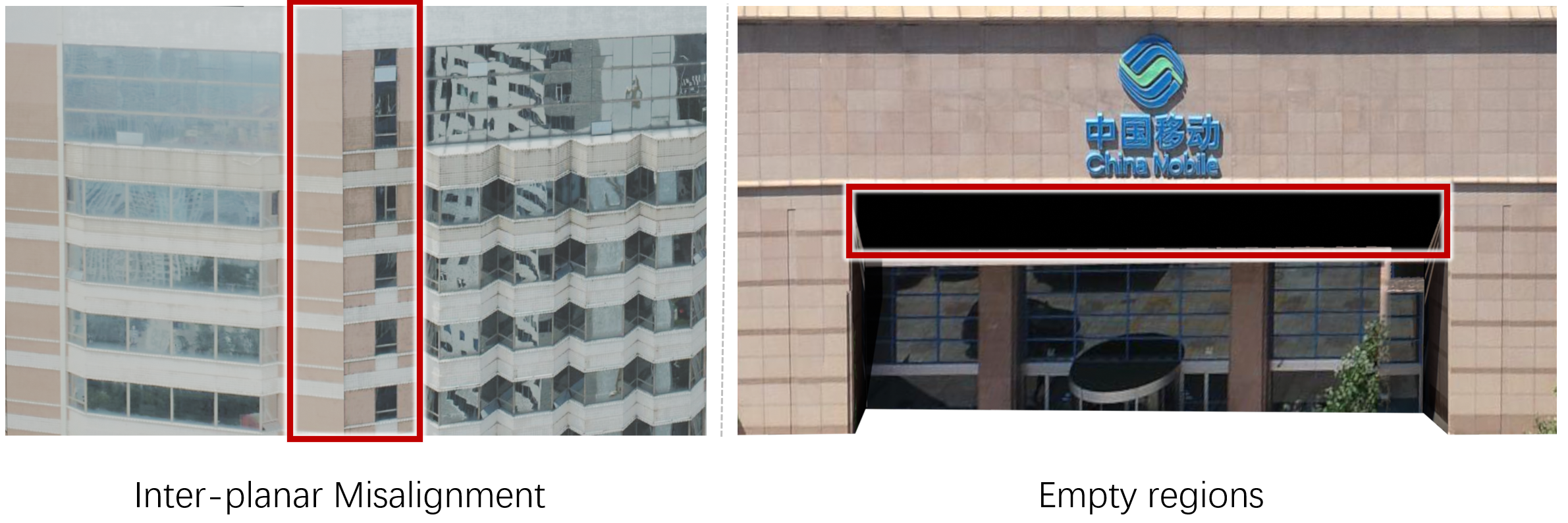}
    \caption{{Failure examples: inter-planar misalignment among two adjacent facades (left), and empty regions covered by none of the input photos (right).}}
	\label{fig:limitations}
\end{figure}

\paragraph{Limitations and future work.}
Our texturing method has several limitations.
First, we process each plane independently, which does not consider the mutual relationship among neighboring planes. 
Fig.~\ref{fig:limitations} (left) shows an example where global alignment of linear features between adjacent planes is not achieved. Note that we here choose a zoomed-in view with the most obvious misalignment among the given buildings to illustrate this issue clearly.
Furthermore, since the texturing of each facade polygon is independent, our plans involve expediting this step with parallel processing.
Second, for model regions covered by none of the input photos, optimization methods are not able to infer the missing region and create content, see Fig.~\ref{fig:limitations} (right).

Finally, significant occlusion and inaccurate camera parameters may affect the final results. In the future, we would like to introduce a pre-processing step to remove the environmental occlusions. Another possible future work can be to remove extrinsic camera needs or jointly optimize camera parameters. Furthermore, since we focus on texture optimization, the camera extrinsic parameters are required as input. Possible future works can be removing such needs for extrinsic cameras as in DUSt3R~\cite{Wang_2024_CVPR}, or jointly optimizing the inaccurate camera parameters.

\section*{Acknowledgments}
We thank the reviewers for their constructive comments. This work was supported in parts by Guangdong S\&T Program (2024B01015004), NSFC (U21B2023, 62302313, 62572059), ICFCRT (W2441020), Guangdong Basic and Applied Basic Research Foundation (2023B1515120026), Shenzhen Science and Technology Program (KJZD20240903100022028, KQTD20210811090044003, RCJC20200714114435012), and Scientific Development Funds from Shenzhen University.

\bibliographystyle{ACM-Reference-Format}
\bibliography{DiffTex}

\end{document}